 \documentclass[final,3p,times,twocolumn]{elsarticle}

\usepackage{amssymb}

\biboptions{sort&compress}

\journal{NIM B}

\begin{document}

\begin{frontmatter}



\title{Activation cross-sections of deuteron induced nuclear reactions on rhenium up to 40 MeV}


\author[1]{F. Ditr\'oi\corref{*}}
\author[1]{F. T\'ark\'anyi}
\author[1]{S. Tak\'acs}
\author[2]{A. Hermanne}
\author[3]{H. Yamazaki} 
\author[3]{M. Baba}
\author[3]{A. Mohammadi} 
\author[4]{A.V.  Ignatyuk}
\cortext[*]{Corresponding author: ditroi@atomki.hu}

\address[1]{Institute of Nuclear Research of the Hungarian Academy of Sciences (ATOMKI),  Debrecen, Hungary}
\address[2]{Cyclotron Laboratory, Vrije Universiteit Brussel (VUB), Brussels, Belgium}
\address[3]{Cyclotron Radioisotope Center (CYRIC), Tohoku University, Sendai, Japan}
\address[4]{Institute of Physics and Power Engineering (IPPE), Obninsk, Russia}

\begin{abstract}
As a part of a thorough work of excitation functions of deuteron induced reactions, experimental cross-sections of $^{185,183m,183g,182}$Os and $^{188,186,184m,184g,183}$Re activation products on $^{nat}Re$ were measured up to 40 MeV for the first time with the activation method using a stacked foil irradiation technique and high resolution -spectrometry. Comparison with the former results of other laboratories and with the predictions of the ALICE-IPPE and EMPIRE-3 model codes, modified for improved calculations for deuteron reactions, and with data in the TENDL-2011 library are also presented. Thick target yields were given deduced from our experimental cross-sections and compared with the few literature values. For practical applications (thin layer activation) also activity versus depth distributions were calculated for selected isotopes.
\end{abstract}


\end{frontmatter}


\section{Introduction}
\label{1}
Integral excitation functions for the production of residual nuclides through light charged particle activation constitutes basic data for different applications. To meet the requirements of different practical applications we started to establish an experimental activation database some years ago by performing new experiments and a systematical survey of existing data of deuteron induced cross-sections up to 50 MeV \cite{1}.
We present here new results on deuteron activation cross-section data on rhenium. No earlier experimental cross-section data were found in the literature. Only a few experimental thick target yields are available at 22 MeV reported by Dmitriev et al. \cite{2}.
Rhenium (atomic mass: 186.207 g/mol, melting point 3180.0 $^{o}C$, density 21.02 $g/cm^{3}$) is a very heavy (dense), high melting point metal. Rhenium is resistant to heat, wear and corrosion. It is able to maintain ductility between absolute zero and its melting point 3180 $^{o}C$. This makes it valuable as an alloying agent for hardening metal components, as well as for use under extreme temperatures.
Rhenium is used in many fields (catalyst, electric components, coatings, etc.). Presently it is mostly used as an additive in super-alloys for aviation technology. Activation cross-sections of proton and deuteron induced reactions are important therefore for estimation of activation dose and for use in thin layer activation technology.
A new application is the production of medical radioisotopes. Some alternate production routes were investigated in detail earlier \cite{3,4,5,6}.

\section{Experimental method and data evaluation}
\label{2}
The experimental method was practically the same as that we applied before for charged particle induced nuclear reactions measurements \cite{7}. Activation method using stacked foils was applied for measuring the whole excitation functions. Two foil-stacks were irradiated at 21 MeV (VUB) and one stack using 40 MeV (CYRIC) deuteron energy. In our experiments high-purity Mo(52.5)Re(47.5) alloy foils (Goodfellow $>99.98 \%$, thickness 50 $\mu m$) were stacked together with Ti monitor foils (10.9 µm) at VUB and Al degrader/monitor foils (100 µm) at CYRIC. The "long" target stack at CYRIC also contained Rh (12.3 $\mu m$), Ho (25 $\mu m$) Au (107 $\mu m$) Yb (25 $\mu m$), Ni(2)Mn(12)Cu(86) alloy (25 $\mu m$) foils for simultaneous measurement of activation cross-sections on these elements. The Ti and Al monitors  foils were used to catch recoiled radioactive products as well as for monitoring the beam intensity and energy by comparing the excitation function of the $^{nat}Ti(d,x)^{48}V$ reaction at VUB and $^{27}Al(d,x)^{22,24}Na$ reactions at CYRIC over the entire energy range. To measure the nominal beam current the stacks were irradiated in a target holder, which served as a Faraday cup and was equipped with collimator (beam diameter on target is 5 mm) and a secondary electron suppressor unit. Irradiations were performed using constant beam current of 160 nA for 60 min (VUB) and 24 nA for 30 min (CYRIC) respectively. 
The number of the produced radionuclides was determined through their decay. For measurement of the activities in the target and the monitor foils HPGe  -ray spectrometers were used. Measurements of the induced activity started in Brussels a few hours after EOB (End of Bombardment), while in Sendai a cooling time of at least a day was set in order to get rid the high activity short-lived isotopes. Chemical separation was not necessary and the measurements in Brussels were performed repeatedly several times (up to months) after EOB, ensuring accurate determination of both short-lived and long-lived activation products (the possibility to measure short-lived activities has been lost in the Sendai experiment). The samples were measured at appropriate detector-sample distances (70 cm to 5 cm) to ensure low dead-times and to decrease pile-ups. The decay- and other nuclear data were collected from the NUDAT2 data-base \cite{8} and are given in Table 1. The Q-values refer to formation of the ground states and taken from \cite{9}. 
The cross-sections were calculated from the well-known activation formula with measured activity, particle flux and number of target nuclei. A number of radionuclides come as a result of cumulative processes, as decay of parent nuclides gives contribution to the production. The particular situation for the individual nuclides will be explained case by case. Elemental rhenium has two stable isotopes with the following abundances: $^{185}Re (37.40\%)$ and $^{187}Re(62.60\%)$. 
The absence of $^{186}Re$ (not stable) in principle allows in many investigated cases and in special energy ranges to obtain pure isotopic cross-sections, required for comparison with model calculations. As in practical applications mostly natural rhenium is used, we have deduced so-called elemental cross-sections, considering the target as monoisotopic.
The number of the target nuclei was determined from the surface density of the foils obtained by precise measurement of the weight and the surface of the metal sheets used before cutting into pieces. The thickness of each individual target and monitor foil was checked for control after cutting.
According to the previous practice the beam current and the incident beam energy were determined by using the excitation function of the re-measured monitor reactions \cite{10, 11}. The energy loss versus depth function in the stack was calculated \cite{12} and refined using the re-measured excitation functions of the monitor reactions as given in \cite{13}.
The latest available decay data were used for our calculations \cite{8}. The cross-section curve of the simultaneously measured monitor reaction is compared with the recommended data as described in our previous works \cite{14}. Acceptably good agreement was found after moderate adjustment on the beam current (10\% compared to the Faraday cup measurement) and on the incident beam energy (0.5 MeV). 
The resulting uncertainty of the cross-sections contains the individual uncertainties of the processes contributing linearly to the final result accordingly to the well-accepted summation rules \cite{15}. The uncertainties of the non-linear processes like half-life, irradiation time, measuring time were neglected in the reported uncertainties of the resulting cross-sections.
The uncertainty of the energy in the targets was estimated from the uncertainty on the energy of the primary beam, the uncertainty in the thickness and uniformity of all foils, the beam straggling and the well-known cumulative effects.

\section{Tehoretical calculations}
\label{3}
In order to avoid large mistakes in evaluations of the measurements, to analyze and understand contributions of individual reactions and to check the predictivity of recently used model codes, the ALICE-IPPE \cite{16}, EMPIRE \cite{17} and TALYS \cite{18} codes are used in our works. While ALICE-IPPE calculates only the total cross-section of the concerned reaction channels, the TALYS and EMPIRE codes permit to calculate a population of different low-lying levels and can thus estimate the isomeric ratios for these levels. Independent data for isomers with ALICE-D code was obtained by using the isomeric ratios calculated with EMPIRE.
During the recent analyses of several deuteron induced reactions we were confronted with poor description of the measured cross-sections due to the high complexity of the interactions. The interaction of deuterons with the target nuclei proceeds largely through direct reaction (DR) processes for deuteron energies below and around the Coulomb barrier, while with increasing incident energy, reaction mechanisms such as pre-equilibrium emission (PE) or evaporation from the fully equilibrated compound nucleus (CN) also become important. The breakup mechanism is responsible for the enhancement of a large variety of reactions along the whole incident-energy range, and thus its contribution to the activation cross-sections has to be explicitly taken into account \cite{19}. There are different possibilities to get better agreement. To get better descriptions in our approach modified codes named ALICE-D and EMPIRE-D were developed at IPPE \cite{14, 20},  in which a simulation of direct (d,p) and (d,t) transitions with general relations for nucleon transfer probability in the continuum \cite{21} is included using an energy dependent enhancement factor for the corresponding transitions \cite{22}. These updated codes were used to analyze the present experimental results. The parameters for the optical model, level densities and pre-equilibrium contributions were taken as described in \cite{23}. In the figures we also present the theoretical results presented in TENDL 2011 data library \cite{24}. The TALYS and TENDL calculations do include a breakup component in all (d,n) and (d,p) reaction channels,  according to our earlier experiences, its strength does not show enough enhancement.

\section{Results}
\label{4}
The new experimental cross-sections are shown in Figs. 1-9 together with results of the theoretical calculations.  We present the results of different irradiations separately to show the possible systematic deviations and errors. The numerical data essential in further data evaluation are collected in Table 2. The main contributing processes are shown in Table 1, together with the reaction Q-values.

\begin{table*}[t]
\tiny
\caption{Decay characteristics of the investigated activation products }
\centering
\begin{center}
\begin{tabular}{l l l l l l} 
\hline 
Nuclide &	Half-life	& E (keV) &	I$_{\gamma}$(\%) &	Contributing reactions	& Q-value (keV)\\
\hline
\shortstack{$^{185}Os$\\ $\epsilon$: 100\%} &	93.6 d &	646.116 & 	78 & 	\shortstack{$^{185}Re(d,2n)$\\ $^{187}Re(d,4n)$} &	\shortstack{-4019.71\\ -17555.9} \\
\hline
\shortstack{$^{183m}Os$\\ IT: 15\%\\ $\epsilon$: 85\%} & 170.715 keV & 	9.9 h	& \shortstack{1034.86\\ 1101.92\\ 1107.9} &	\shortstack{6.02\\ 49.0\\ 22.4}	\shortstack{$^{185}Re(d,4n)$\\ $^{187}Re(d,6n)$} &	\shortstack{-19479.6\\ -33015.8}\\
\hline
\shortstack{$^{183g}Os$\\ $\epsilon$: 100\%} &	\shortstack{13.0 h\\ \ \ \ }	& \shortstack{114.47\\  381.763} &	\shortstack{20.6\\ 89.6}	& \shortstack{$^{185}Re(d,4n)$\\ $^{187}Re(d,6n)$} & \shortstack{-19308.9\\ -32845.1}\\
\hline
\shortstack{$^{182}Os$\\ $\epsilon$: 100\%} & \shortstack{21.84 h\\ \ \ } &	\shortstack{130.80\\ 180.20} &	\shortstack{3.30\\ 34.1} & \shortstack{$^{185}Re(d,5n)$\\ $^{187}Re(d,7n)$} &	\shortstack{-26433.9\\ -39970.1}\\
\hline
\shortstack{$^{188}Re$\\ $\beta^{-}$: 100\%} &	17.004 h &	155.041 & 	15.61 & 	$^{187}Re(d,p)$ &	3647.184\\
\hline
\shortstack{$^{186}Re$\\ $\beta^{-}$: 92.53\%\\ $\epsilon$: 7.47\%} &	3.7183 d &	137.157	& 9.47	& \shortstack{$^{185}Re(d,p)$\\ $^{187}Re(d,p2n)$}	& \shortstack{3954.794\\ -9581.4}\\
\hline
\shortstack{$^{184m}Re$\\ IT:74.5\%\\ $\epsilon$: 25.5\%\\ 188.0463 keV} &	169 d &	\shortstack{104.7395\\ 161.269\\ 216.547\\ 252.845} & \shortstack{13.6\\ 6.56\\ 9.5\\ 10.8} & \shortstack{$^{185}Re(d,p2n)$\\ $^{187}Re(d,p4n)$} & \shortstack{-10079.4\\ -23545.6}\\
\hline
\shortstack{$^{184g}Re$\\ $\epsilon$: 100\%} &	35.4 d	& \shortstack{111.2174\\ 792.067\\ 894.760\\ 903.282} & \shortstack{17.2\\ 37.7\\ 15.7\\ 38.1} & \shortstack{$^{185}Re(d,p2n)$\\ $^{187}Re(d,p4n)$} & \shortstack{-9891.4\\ -23427.59}\\
\hline
\shortstack{$^{183}Re$\\ $\epsilon$: 100\%} &	70.0 d & \shortstack{162.3266\\ 291.7282} & \shortstack{23.3\\ 3.05} & \shortstack{$^{185}Re(d,p3n)$\\ $^{187}Re(d,p5n)$} & \shortstack{-16378.32\\ -29914.52}\\
\hline
\shortstack{$^{182m}Re$\\ $\epsilon$: 100\%\\ 0+X keV} & 12.7 h	& \shortstack{100.12\\ 152.43\\ 470.26*\\ 894.85*\\ 1121.4\\ 1189.2\\ 1221.5}	& \shortstack{14.4\\ 7.0\\ 2.02\\ 2.11\\ 32.0\\ 15.1\\ 25.0} & \shortstack{$^{185}Re(d,p4n)$\\ $^{187}Re(d,p6n)$} &	\shortstack{-24813.0+\\ -38349.0+}\\
\hline
\shortstack{$^{182g}Re$\\ $\epsilon$: 100\%} &	64.0 h &	\shortstack{100.10\\ 130.81*\\ 169.15*\\ 191.39*\\ 229.32\\ 286.56*\\ 351.07*\\ 1076.2*\\ 1121.3\\ 1221.4\\ 1231.0\\ 1427.3*} & \shortstack{16.5\\ 7.5\\ 11.4\\ 6.7\\ 25.8\\ 7.1\\ 10.3\\ 10.6\\ 22.1\\ 17.5\\ 14.9\\ 9.8} & \shortstack{$^{185}Re(d,p4n)$\\ $^{187}Re(d,p6n)$} & \shortstack{-24813.0\\ -38349.0}\\
\hline

\end{tabular}
\end{center}
\begin{flushleft}
\footnotesize{+ The energy of the excited state is unknown\\
When complex particles are emitted instead of individual protons and neutrons the Q-values have to be decreased by the respective binding energies of the compound particles: np-d, +2.2 MeV; 2np-t, +8.48 MeV; n2p-$^3$He, +7.72 MeV; 2n2p- , +28.30 MeV\\
The independent $\gamma$-lines are marked with *}
\end{flushleft}

\end{table*}

\subsection{Cross-sections}
\label{4.1}

\subsubsection{Production of osmium isotopes}
\label{4.1.1}
The measured excitation functions came in all cases from a combination of (d,xn) reactions on the $^{185}Re$ and $^{187}Re$.

\textbf{$^{nat}Re(d,x)^{185}Os$}

The main contributing processes are the direct production via the $^{185}Re(d,2n)$ and $^{187}Re(d,4n)$ reactions. The experimental and theoretical excitation functions are shown in Fig. 1. The results of different irradiations show acceptably good agreement in the overlapping energy region. The theoretical predictions differ by factor of two.

\begin{figure}[h]
\includegraphics[scale=0.3]{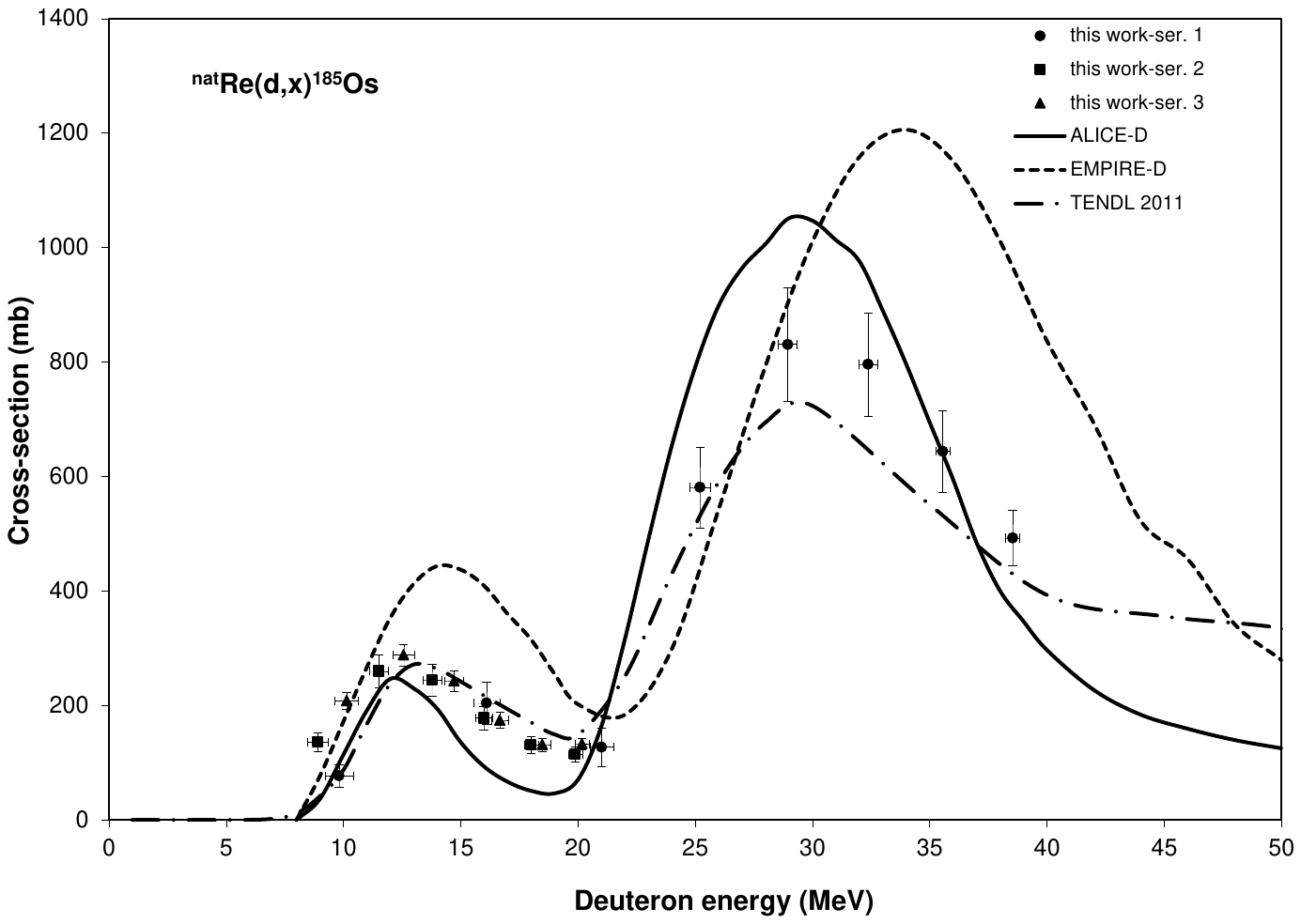}
\caption{Experimental and theoretical excitation functions for $^{nat}Re(d,x)^{185}Os$}
\end{figure}

\textbf{$^{nat}Re(d,x)^{183m}Os$ and $^{nat}Re(d,x)^{183g}Os$}

The Q-value of the two contributing reactions $^{185}Re(d,4n)$ and $^{187}Re(d,6n)$ are collected in Table 1. We present independent cross-sections for production of the metastable state (Fig. 2) and for the ground state (Fig. 3) obtained after a small correction (low  $\sigma_m$, IT=15 \%) from the decay of metastable state. Theories in average reproduce the shape and the magnitude of the experimental data, but in the investigated energy range there are large differences in the magnitude of the different model predictions.

\begin{figure}[h]
\includegraphics[scale=0.31]{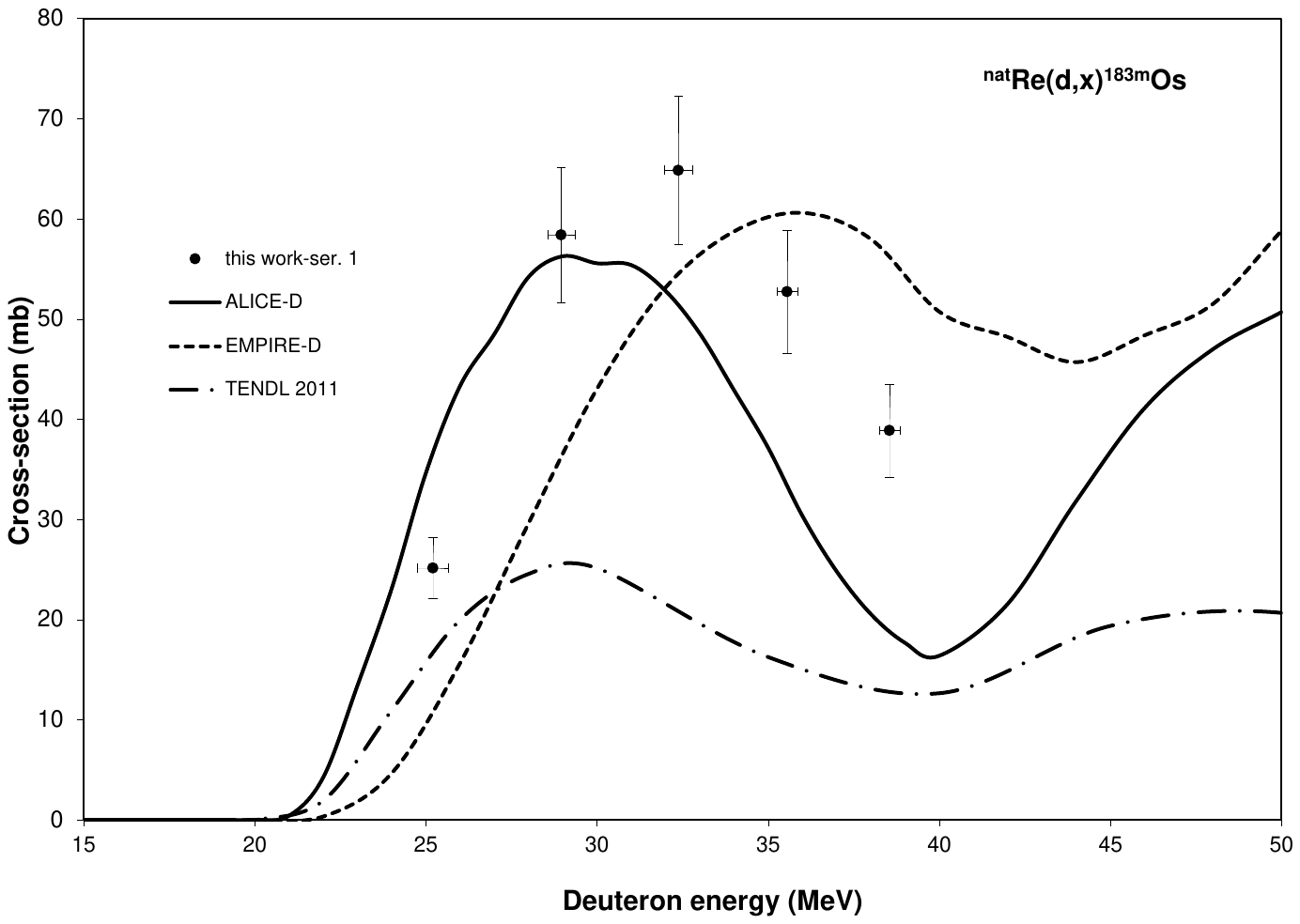}
\caption{Experimental and theoretical excitation functions for $^{nat}Re(d,x)^{183m}Os$}
\end{figure}

\begin{figure}[h]
\includegraphics[scale=0.31]{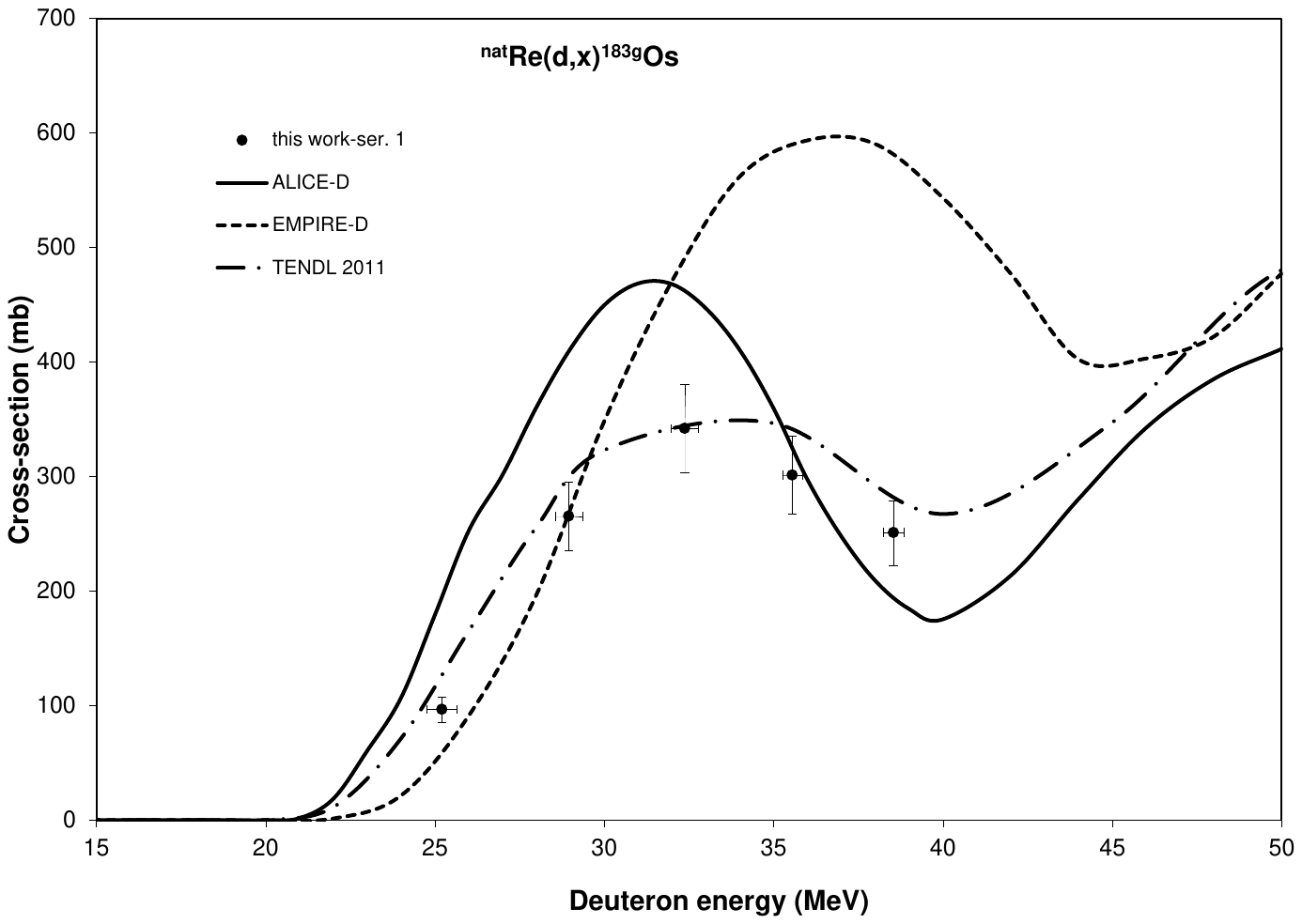}
\caption{Experimental and theoretical excitation functions for $^{nat}Re(d,x)^{183g}Os$}
\end{figure}

$^{nat}Re(d,x)^{182}Os$

Only in a few foils in the beginning of the stack we can detect the  $\gamma$-lines from the decay of the ${182}Os$. The few obtained experimental data points fit well to the theoretical excitation functions (see Fig. 4). In the investigated energy range only the $^{185}Re(d,5n)$ contribute to the production of $^{182}Os$.

\begin{figure}[h]
\includegraphics[scale=0.31]{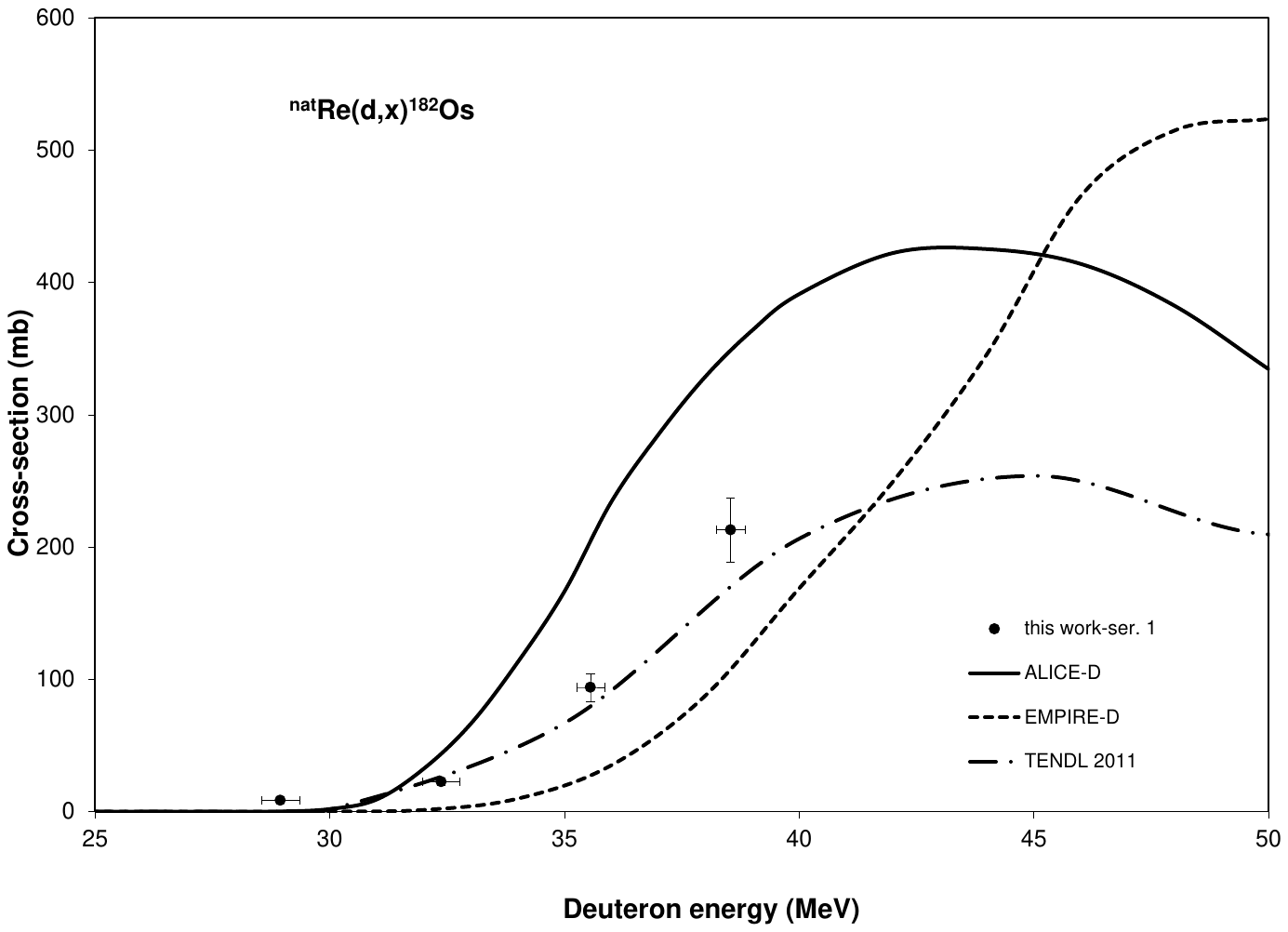}
\caption{Experimental and theoretical excitation functions for $^{nat}Re(d,x)^{182}Os$}
\end{figure}

\subsubsection{Production of rhenium isotopes}
\label{4.1.2}

The measured excitation functions came in all cases from a combination of (d,xn) reactions on the $^{185}Re$ and $^{187}Re$. The level schemes for $^{182}Re$ and $^{184}Re$ are estimated with large uncertainties for some important  $\gamma$-transitions. So, an accuracy of calculations for the corresponding isomer yields cannot be very high.

$^{nat}Re(d,x)^{188}Re$

The $^{188}Re$ is produced only through $^{187}Re(d,p)$ reaction. The $^{188}Re$ could also be produced from $^{187}Re$ through (n,$\gamma$ ) reaction by neutrons coming from deuteron break-up. This effect was excluded, because out of the target foils referred in the experimental part one more foil of the same material was inserted at the end of the stack beyond the threshold of the $^{nat}Re(d,x)^{188}Re$ reaction and it did not contain $^{188}Re$. It means that the neutron induced production is negligible.  The measured and calculated excitation functions are shown in Fig. 5. There is a strong underestimation of the measured cross-sections in case of TENDL 2011. The agreement with the maximum value of the updated version ALICE-D and EMPIR-D is better, but still with underestimation.  In case of EMPIRE-D the reproduction of the shape of the high energy tail of experimental excitation function is also poor. 

\begin{figure}[h]
\includegraphics[scale=0.31]{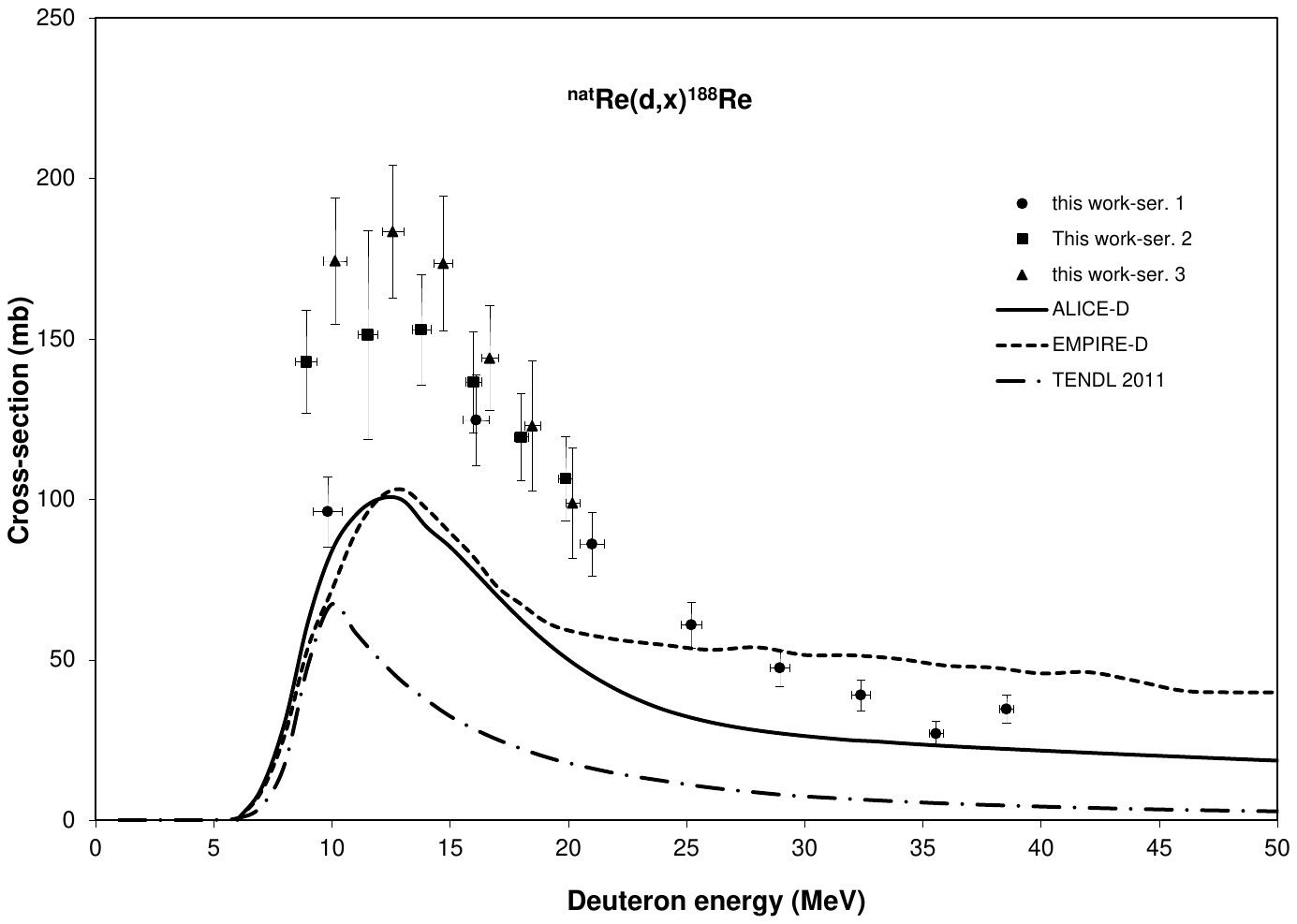}
\caption{Experimental and theoretical excitation functions for $^{nat}Re(d,x)^{188}Re$}
\end{figure}

$^{nat}Re(d,x)^{186g}Re$

The $^{186g}Re$ is produced through the $^{185}Re(d,p)$ and $^{187}Re(d,p2n)$ reactions. As discussed in in the previous section the neutron induced production is also negligible here because of the lack of $^{186g}Re$ in the last foil. There is a strong underestimation of the measured cross-sections in case of all models for (d,p) reaction (Fig. 6). The situation is the same in the case of (d,p2n) for ALICE-D, EMPIRE-D and TALYS (TENDL-2011).

\begin{figure}[h]
\includegraphics[scale=0.31]{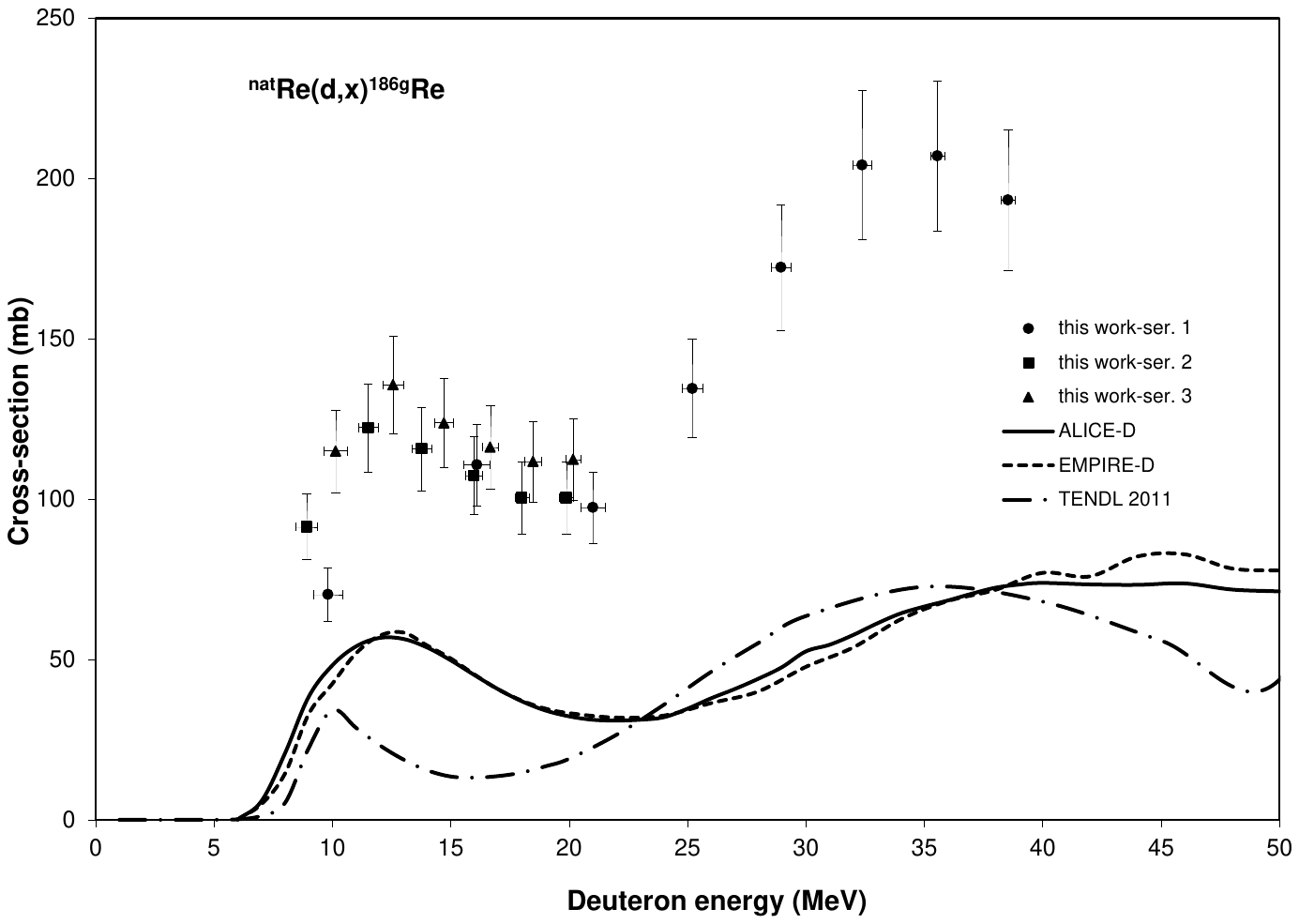}
\caption{Experimental and theoretical excitation functions for $^{nat}Re(d,x)^{186g}Re$}
\end{figure}

$^{nat}Re(d,x)^{184m}Re$ and $^{nat}Re(d,x)^{184g}Re$

The experimental and theoretical cross-sections are shown in Figs. 7-8. The models describe well the shape, but the estimation of the magnitude is not so successful, probably due to the unclear level scheme.

\begin{figure}[h]
\includegraphics[scale=0.31]{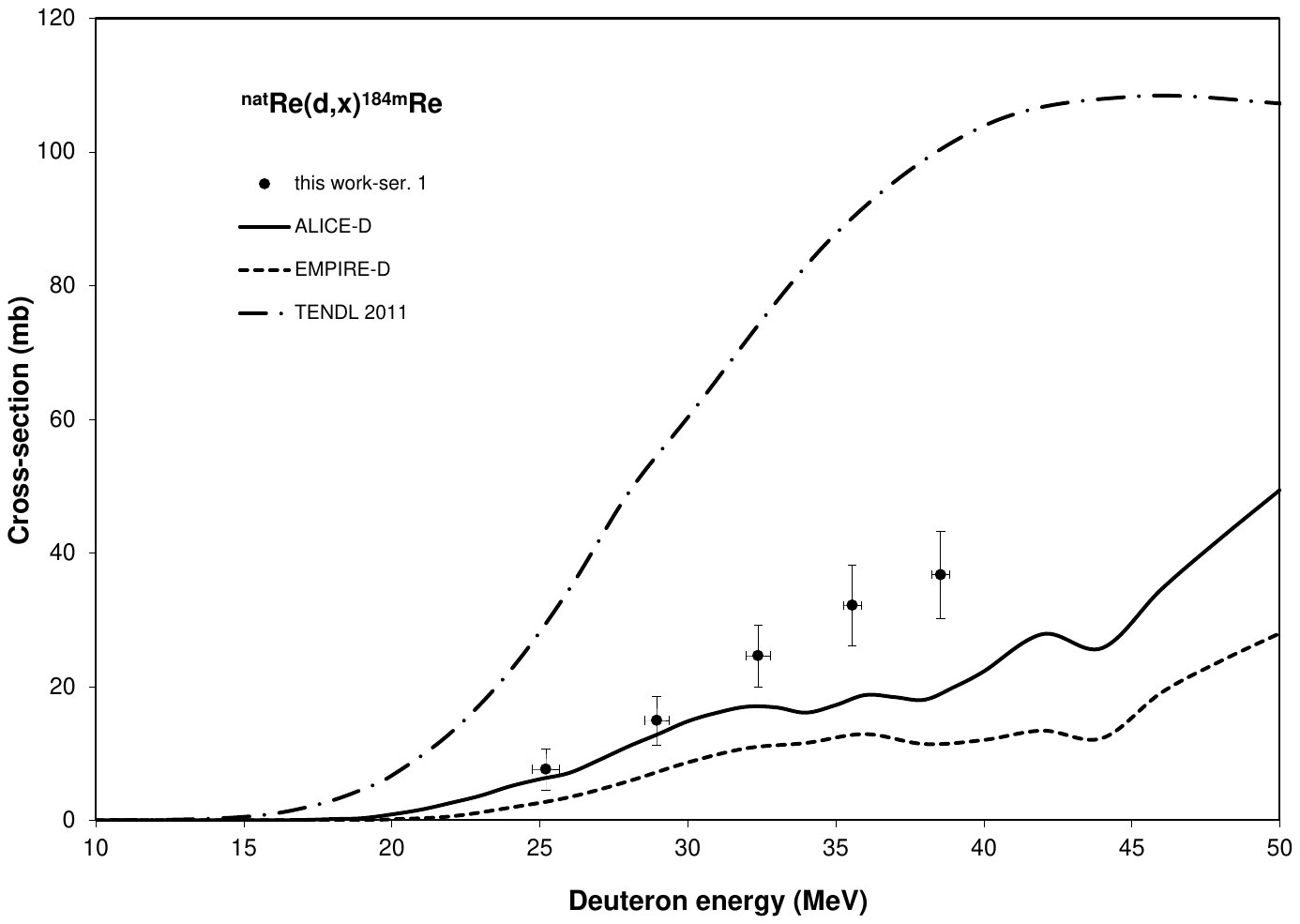}
\caption{Experimental and theoretical excitation functions for $^{nat}Re(d,x)^{184m}Re$}
\end{figure}

\begin{figure}[h]
\includegraphics[scale=0.31]{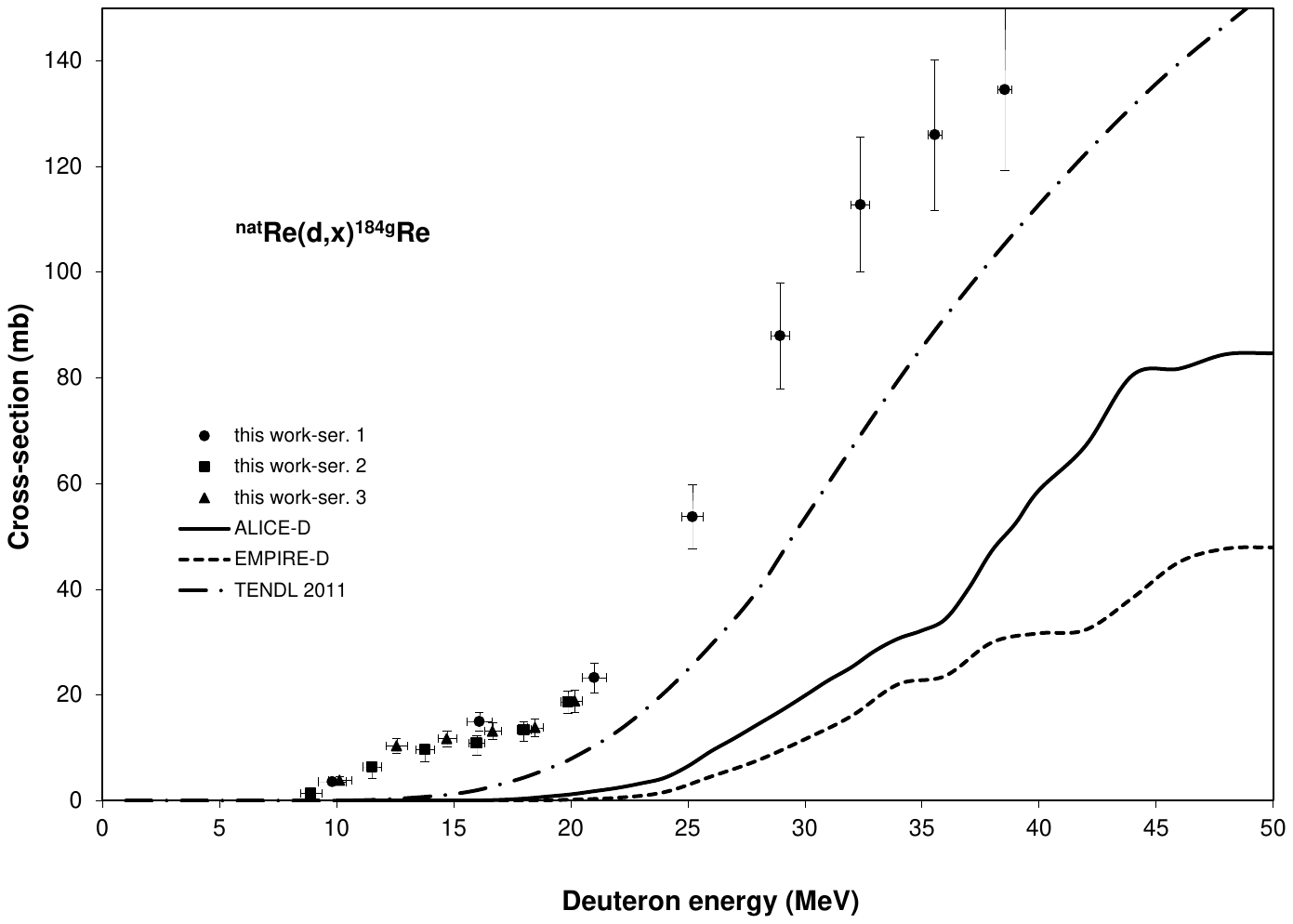}
\caption{Experimental and theoretical excitation functions for $^{nat}Re(d,x)^{184g}Re$}
\end{figure}

$^{nat}Re(d,x)^{183}Re$ (cum)

The $^{183}Re$ ($T_{1/2} = 70.0 d$) is produced directly via (d,pxn) reactions ($^{185}Re(d,p3n) + 
^{187}Re(d,p5n)$ and by decay of the $^{183m}Os$ ($\epsilon$:85\%, $T_{1/2}=9.9 h$), and $^{183g}Os$($\epsilon$:100\%, $T_{1/2} = 13.0 h$). Our cross-sections were deduced after complete decay of the parent isomers, in such a way they represent cumulative cross-sections. The measured and calculated cross-sections are shown in Fig. 9. The ALICE-D, EMPIRE-D and TENDL-2011 describe approximately the shape of the experimental excitation function. In magnitude the best agreement is with ALICE-D.

\begin{figure}[h]
\includegraphics[scale=0.31]{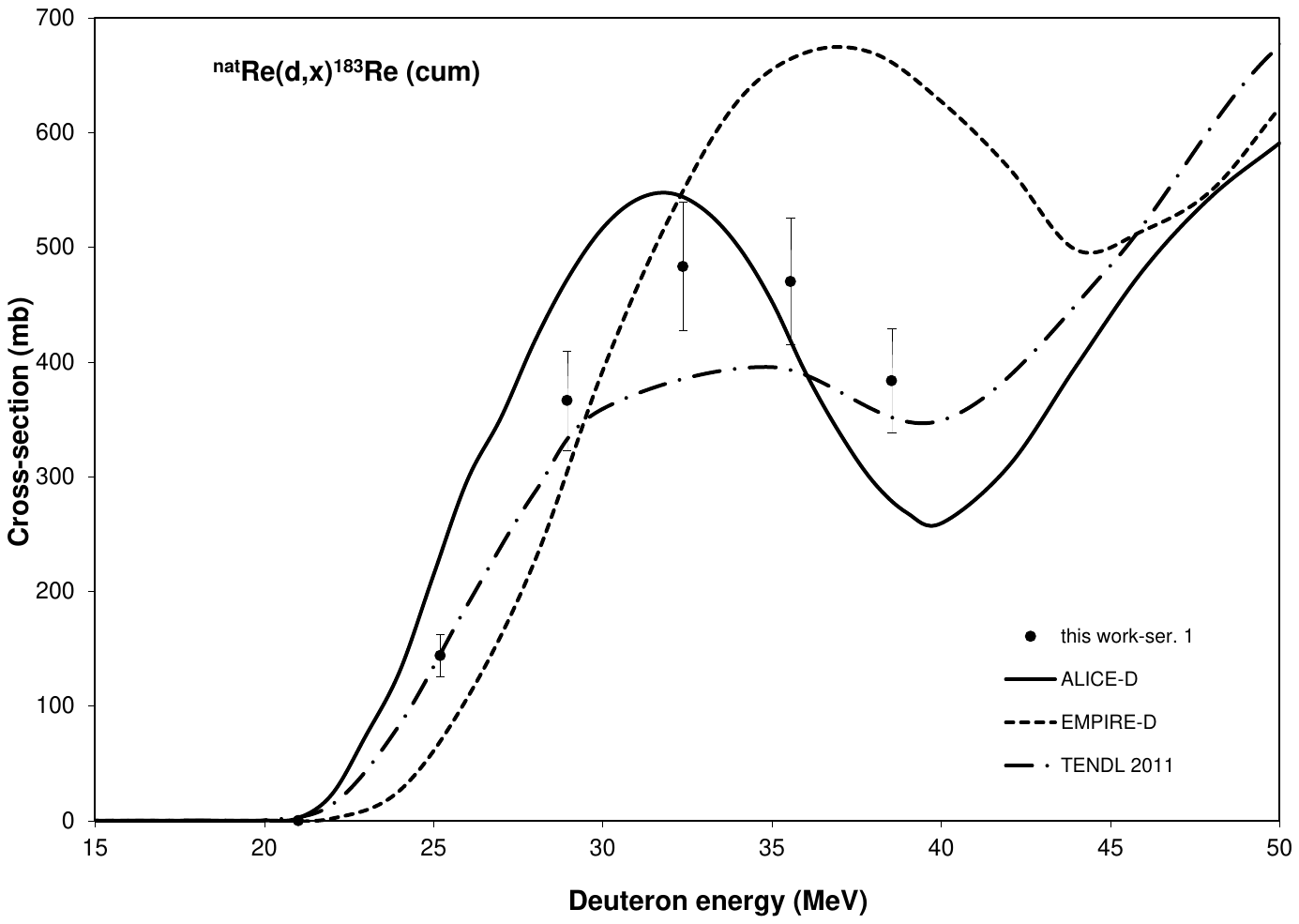}
\caption{Experimental and theoretical excitation functions for $^{nat}Re(d,x)^{183}Re$ (cum)}
\end{figure}

$^{nat}Re(d,x)^{182m}Re$ and ${nat}Re(d,x)^{182g}Re$

The $^{182m}Re$ ($T_{1/2} = 12.7 h$) is produced both directly and by the decay of the simultaneously produced $^{182}Os$. The $^{182}Os$ ($T_{1/2} = 21.84 h$) is decaying only to $^{182m}Re$ ($T_{1/2} = 12.7 h$).  From other side the $^{182g}Re$ (64 h) is only produced by direct production route (no decay contribution from $^{182}Os$ and no internal transition from the isomer. According to the theoretical estimations the cross-sections for the direct production of both isomeric states are low in the investigated energy range and the Q value of $^{185}Re(d,p4n)$ reaction is high (-24.8 MeV). 
As it was mentioned, the $^{182m}Re$ is produced also through the decay of the significantly higher cross-section $^{182}Os$. The $^{182m}Re$ has strong common  $\gamma$-lines with the $^{182g}Re$ (see Table 1.) and a few independent weak  $\gamma$-lines (470.2 keV and 894.85 keV). We could detect these independent  $\gamma$-lines in our spectra only with poor statistics. We have recognized that the separation of the high yield contribution of the parent could be done only with very large uncertainties and so with low reliability. 
Therefore we have tried to find direct production cross-section of the 182gRe. The $^{182g}Re$ has a few moderately strong independent  $\gamma$-lines: 130.81 keV(7.5\%), 169.15 keV(11.4\%), 191.39 keV(6.7\%), 286.56 keV(7.1\%), 351.07 keV(10.3\%) and 1076.2 keV(10.6\%). Unfortunately due to the complex  $\gamma$-spectra caused by numerous reactions on Mo and Re target isotopes, the reliable identification of these  $\gamma$-lines was unsuccessful. The theoretical codes give also very discrepant results for these two reactions.

\begin{table*}[t]
\tiny
\caption{Measured cross-sections for the Os and Re radioisotopes }
\centering
\begin{center}
\begin{tabular}{|r| r| r| r| r| r| r| r| r| r| r| r| r| r| r| r| r| r| r| r| r|} 
\hline 
series &	\multicolumn{2}{|c|}{Bombarding}	&	\multicolumn{2}{|c|}{$^{185}Os$}&		\multicolumn{2}{|c|}{$^{183m}Os$} &		\multicolumn{2}{|c|}{$^{183g}Os$} &		\multicolumn{2}{|c|}{$^{182}Os$} &		\multicolumn{2}{|c|}{$^{188}Re$} &		\multicolumn{2}{|c|}{$^{186}Re$} &	\multicolumn{2}{|c|}{$^{184m}Re$} &		\multicolumn{2}{|c|}{$^{184g}Re$} &		\multicolumn{2}{|c|}{$^{183}Re$(cum)}\\
\cline{2-21}

 & E	& $\pm\delta E$ & $\sigma$ & 	$\pm\delta\sigma$ & $\sigma$ & 	$\pm\delta\sigma$& $\sigma$ & 	$\pm\delta\sigma$& $\sigma$ & 	$\pm\delta\sigma$& $\sigma$ & 	$\pm\delta\sigma$& $\sigma$ & 	$\pm\delta\sigma$& $\sigma$ & 	$\pm\delta\sigma$& $\sigma$ & 	$\pm\delta\sigma$& $\sigma$ & 	$\pm\delta\sigma$\\
\hline
 & \multicolumn{2}{|c|}{MeV} & \multicolumn{2}{|c|}{mb}& \multicolumn{2}{|c|}{mb}& \multicolumn{2}{|c|}{mb}& \multicolumn{2}{|c|}{mb}& \multicolumn{2}{|c|}{mb}& \multicolumn{2}{|c|}{mb}& \multicolumn{2}{|c|}{mb}& \multicolumn{2}{|c|}{mb}& \multicolumn{2}{|c|}{mb}\\
 \hline
1	& 38.5	& 0.3	& 492.58	& 56.49	& 38.86	& 4.61	& 251.0	& 28.3	& 213.0	& 24.0	& 34.6	& 4.4	& 193.3	& 22.0	& 36.7	& 6.5	&  134.5	& 15.3	& 383.7	& 45.3\\
&	35.6&	0.3&	644.06&	73.50&	52.72&	6.14&	301.3&	33.9&	94.0&	10.7&	27.0&	3.9&	207.1&	23.5&	32.2&	6.0&	126.0&	14.3&	470.3&	55.2\\
&	32.4&	0.4&	796.01&	90.15&	64.84&	7.44&	342.0&	38.5&	22.6&	2.8&	39.0&	4.8&	204.2&	23.2&	24.6&	4.7&	112.8&	12.8&	483.4&	56.0\\
&	29.0&	0.4&	830.75&	94.08&	58.40&	6.73&	265.4&	29.9&	8.6&	1.4&	47.4&	5.7&	172.3&	19.6&	14.9&	3.7&	88.0&	10.0&	366.4&	43.2\\
&	25.2&	0.5&	581.05&	66.11&	25.15&	3.02&	96.8&	11.0& & &		 	60.9&	7.1&	134.6&	15.4&	7.6&	3.1&	53.7&	6.1&	144.0&	18.6\\
&	21.0&	0.5&	127.24&	16.29&&&&&&&		 		 		  	86.1&	 9.9&	 97.4&	11.2&	&&	 	23.3&	2.8&&\\
&	16.1&	0.6&	204.27&	24.76&&&&&&&		 		 		 	124.7&	14.2&	110.8&	12.7&	&&	 	15.0&	1.8&&\\
&	9.8&	0.6&	77.13&	10.22&&&&&&&	 	 	 	 	 	 	  96.1&	10.9&	 70.3&	8.2&	&& 	 	 3.6&	0.6&&\\
\hline
2&	19.9&	0.3&	114.60&	13.17&&&&&&&		 		 		 	106.4&	13.2&	100.5&	11.3&	&&	 	18.7&	2.1&&\\
&	18.0&	0.3&	131.14&	15.00&&&&&&&		 		 		 	119.4&	13.6&	100.5&	11.3&	&&	 	13.4&	1.6&&\\
&	16.0&	0.4&	178.15&	20.67&&&&&&&		 		 		 	136.5&	15.8&	107.4&	12.1&	&&	 	10.9&	1.4&&\\
&	13.8&	0.4&	244.52&	27.75&&&&&&&		 		 		 	152.9&	17.2&	115.7&	13.0&	&&	 	9.6&	1.2&&\\
&	11.5&	0.4&	260.05&	29.32&&&&&&&		 		 		 	151.3&	32.5&	122.3&	13.7&	&&	 	6.4&	0.8&&\\
&	8.9&	0.5&	136.23&	15.95&&&&&&&	 	 	 	 	 	 	142.8&	16.1&	 91.4&	10.3&	&& 	 	1.3&	0.5&&\\
\hline
3&	20.2&	0.3&	61.15&	9.13&&&&&&&		 		 		 	   98.8&	17.4&	112.4&	12.6&	&&	 	18.9&	2.2&&\\
&	18.5&	0.4&	60.23&	11.37&&&&&&&		 		 		 	122.9&	20.3&	111.8&	12.6&	&&	 	13.8&	1.7&&\\
&	16.7&	0.4&	89.06&	13.57&&&&&&&		 		 		 	144.0&	16.4&	116.3&	13.1&	&&	 	13.1&	1.6&&\\
&	14.7&	0.4&	128.14&	17.73&&&&&&&		 		 		 	173.5&	21.0&	123.8&	13.9&	&&	 	11.7&	1.5&&\\
&	12.6&	0.5&	144.59&	19.10&&&&&&&		 		 		 	183.4&	20.6&	135.7&	15.2&	&&	 	10.4&	1.4&&\\
&	10.1&	0.5&	100.60&	14.22&&&&&&&		 		 		 	174.3&	19.6&	115.0&	12.9&	&&	 	3.8&	0.8&&\\
\hline

\end{tabular}
\end{center}
\end{table*}

\section{Thick target yields and activity distribution}
\label{5}

In different practical applications either the total produced activity or the distribution of that activity as a function of bombarding energy and/or the depth are often more convenient to use than the microscopic cross-section data. The production yield is directly connected to the cross-section and to the composition of the used target.
From fits to our experimental excitation functions thick target yields were calculated and are shown in Fig 10 in comparison with the experimental thick target yield data in the literature. The deduced yields are so called physical yields, calculated for an instantaneous irradiation \cite{25}. Experimental thick target yields were found only for $^{185}Os$ \cite{26} and $^{188}Re$ \cite{27} in the literature and are in good agreement with our calculated results.
Some of the investigated radionuclides (depending on half-life, excitation functions, emitted  $\gamma$-lines) produced in deuteron bombardment on natural rhenium are suitable for thin layer activation (TLA) applications to investigate the wear or corrosion rate of friction surfaces. Due to the short half-life the 186Re (3.72 d) radiotracer can be used for monitoring processes with quick surface loss, while the $^{185}Os$ (93.6 d) is suitable for investigating long term slow processes. The activation curves at optimized bombarding energies for both reactions (15 MeV for 186Re and 13.8 MeV for $^{185}Os$) and 1 h and  A irradiation and also for 10 d cooling time are shown in Fig. 11.

\begin{figure}[h]
\includegraphics[scale=0.31]{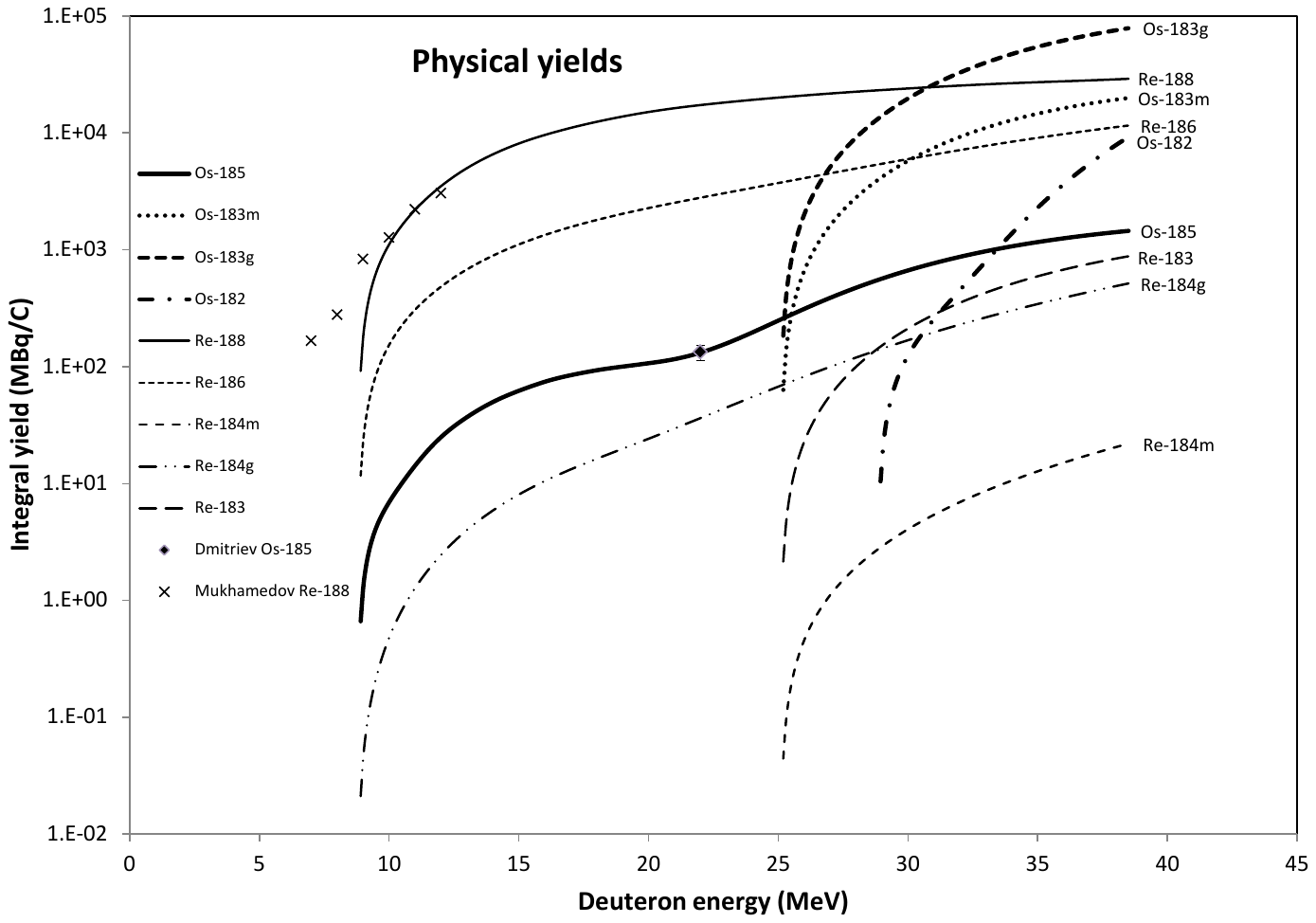}
\caption{Thick target yields calculated from excitation functions for production of $^{185,183m,183g,182}Os$ and $^{188,186,184m,184g,183}Re$}
\end{figure}

\begin{figure}[h]
\includegraphics[scale=0.31]{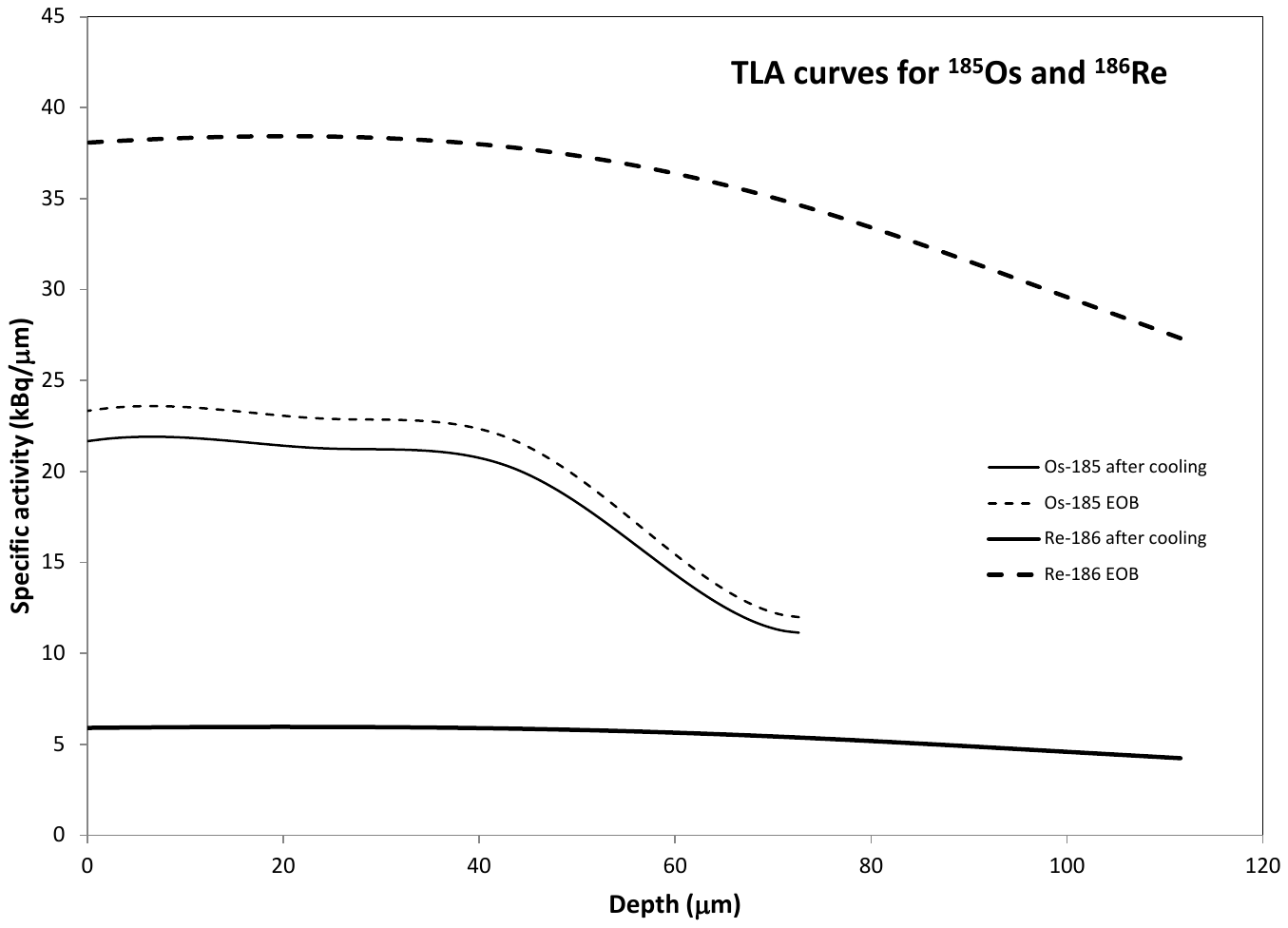}
\caption{The $^{186}Re$ and $^{188}Os$ activation curves with 15 and 13.8 MeV bombarding energies respectively (beam current: 1  $\mu$A; irradiation time 1 h; cooling time: 10 d)}
\end{figure}

\section{Summary and conclusions}
\label{6}
 
The principal aim of this investigation was twofold: in the frame of a systematic study of activation cross-sections of the deuteron induced reactions of metals to measure the missing activation data on rhenium and to check the predictivity of the most widely used theoretical codes. 
We present for the first time excitation curves for deuteron induced reactions on natRe targets up to 40 MeV leading to the production of $^{185,183m,183g,182}Os$ and $^{188,186,184m,184g,183}Re$ radionuclides. The results are correlating to measurements of monitor reaction over the whole energy range. The results of irradiations of different primary beam energy show good agreement in the overlapping energy ranges.
Comparison of different model codes and data libraries with experiments showed modest agreement, even the updated versions (modified for deuteron induced reactions) of the codes ALICE and EMPIRE are not always convincing. Although the inclusion of a phenomenological enhancement factor for direct (d,p) reactions and explicit description of more compound exit channels has improved the description in some cases, still very large discrepancies remain in other cases. In average the agreement between the experimental data and the TALYS theoretical results in TENDL-2011 is also not satisfactory. 
These new results can have applications in accelerator technology, radiation protection and nuclear reaction theory development. The activation cross-section data of proton induced reactions in rhenium are also missing. The data evaluation of the proton experiments is in progress.

\section{Acknowledgements}
\label{7}

This study was performed in the frame of the MTA-JSPS and MTA-FWO (Vlaanderen) collaboration programs. The authors thank the different research projects and their respective institutions for the practical help and providing the use of the facilities for this study.
 



\clearpage
\bibliographystyle{elsarticle-num}
\bibliography{Red}







\end{document}